# Exchange bias associated with phase separation in the Nd$_{2/3}$Ca$_{1/3}$MnO$_3$ manganite

Elena Fertman, Sergiy Dolya, and Vladimir Desnenko,

B. Verkin Institute for Low Temperature Physics and Engineering NASU, 47 Lenin Ave., Kharkov 61103, Ukraine

email address: fertman@ilt.kharkov.ua

Marcela Kajňaková and Alexander Feher

Institute of Physics, Faculty of Science, P. J. Šafárik University in Košice, Faculty of Science, Park Angelinum 9, 04154 Košice, Slovakia

**Abstract**

The exchange bias (EB) phenomenon has been found in Nd$_{2/3}$Ca$_{1/3}$MnO$_3$ perovskite. The phenomenon manifests itself as a negative horizontal shift of magnetization hysteresis loops. The EB phenomenon is evident of an interface exchange coupling between coexisting antiferromagnetic (AFM) and ferromagnetic (FM) phases and confirms the phase separated state of the compound at low temperatures. The EB effect is found to be strongly dependent on the cooling magnetic field and the temperature, which is associated with the evolution of spontaneous AFM – FM phase separated state of the compound. Analysis of magnetic hysteresis loops has shown that ferromagnetic moment $M_{FM}$ originating from the FM clusters saturates in a relatively low magnetic field about $H \sim 0.4$ T. The obtained saturation value $M_{FM}$ (1 T) $\sim 0.45$ $\mu_B$ is in a good agreement with our previous neutron diffraction data.

# 1. Introduction

Exchange bias phenomena is a property of coupled antiferromagnetic (AFM) – ferromagnetic (FM) systems that occurs due to magnetic interface effects [1 – 3]. It manifests itself as a shift ($H_{EB}$) of the magnetization hysteresis loop along the field axis when the system is cooled down in an external magnetic field through the magnetic ordering temperatures of the both AFM and FM phases. The EB phenomena is observed in many different systems containing FM – AFM interfaces, such as small particles [4,5], thin films [6,7], martensitic alloys [8], nanostructures [9]. So far research on exchange bias has been mainly focused on artificially made systems, like FM-AFM multilayers, where a well defined and controllable interface exist [2, 10]. Not so long ago the first evidence of the EB effect was found in spontaneously phase-separated manganite $Pr_{2/3}Ca_{2/3}MnO_3$, where FM nanodomains are immersed within a charge ordered AFM host [11]. This finding has shown that in a spontaneously phase-separated systems the exchange coupling at the interfaces between the FM regions and the surrounding AFM matrix may create an unidirectional anisotropy [3]. Soon the EB effect was observed in other phase separated perovskite manganites $Y_{0.2}Ca_{0.8}MnO_3$ [12], $La_{0.5}Sr_{0.5}MnO_3$ [13], $Pr_{0.5}Ca_{0.5}MnO_3$ [14] and cobaltites [15 – 18].

In this paper, we report the observation of exchange bias phenomenon in spontaneously phase-separated colossal magnetoresistance compound perovskite $Nd_{2/3}Ca_{1/3}MnO_3$, in which intrinsic phase inhomogeneity plays a crucial role [19,20]. The compound is a single phased at room temperature. Lattice, charge, orbital, and spin degrees of freedom are closely related in the compound. Below the room temperature the compound exhibits a sequence of phase transformations leading to its magnetic phase segregated ground state which represents nano-sized ferromagnetic clusters immersed in the charge-ordered antiferromagnetic matrix.

The charge ordering phase transformation in $Nd_{2/3}Ca_{1/3}MnO_3$, which takes place at $T_{CO} \approx 212$ K [21], is of the first order martensitic type. It leads to the self-organized coexistence of charge-ordered (CO) and charge-disordered (CD) phases in a wide temperature range below the room temperature. Extended temperature hysteresis of the magnetic susceptibility in the charge ordering region is one of the evidences of the nonequilibrium phase-segregated state. Below $T_{CO}$ the compound exhibits a sequence of magnetic transformations: two antiferromagnetic ones at $T_{N1} \sim 130$ K and $T_{N2} \sim 80$ K, and a ferromagnetic one at $T_C \sim 70$ K [22,23]. They lead to the coexistence of at least three different magnetic phases at low temperatures: two AFM charge-ordered ones and the FM charge-disordered one. The low temperature magnetic phase-segregated state arises from the CO–CD phase segregated state: CO phase becomes antiferromagnetic, CD phase becomes ferromagnetic. Magnetic behavior of the $Nd_{2/3}Ca_{1/3}MnO_3$ is consistent with a cluster-glass magnetic state below the freezing temperature $T_g \sim 60$ K, which is close to the Curie temperature $T_C \sim 70$ K [19].

In the present study we have observed the exchange bias phenomena in the $Nd_{2/3}Ca_{1/3}MnO_3$ compound when it is cooled down in a static magnetic field below $T_C$. The effect is revealed to be strongly dependent on the cooling magnetic field and temperature. The exchange bias found confirms the phase segregated state of $Nd_{2/3}Ca_{1/3}MnO_3$ compound and get some insight into the properties of the spontaneous phase-segregated exchange biased system.

## 2. Experiment

A polycrystalline $Nd_{2/3}Ca_{1/3}MnO_3$ compound was prepared by a standard solid state reaction technique from stoichiometric amounts of proper powders. X-ray crystal-structure analysis indicated a single-phase material (space group, *Pnma*) at room temperature.

Magnetic measurements were made using Quantum Design Magnetic Properties Measurement System (MPMS) and a noncommercial superconducting quantum interference device (SQUID) magnetometer. Magnetic hysteresis loops were measured at 10 - 80 K after cooling the sample in zero magnetic field and after cooling in an applied fields $H_{cool}$ up to 5 T. After each *M(H)* hysteresis loop measuring the sample was demagnetized by warming up to 320 K and then by delay at 320 K for 1800 sec.

## 3. Results and discussion

When $Nd_{2/3}Ca_{1/3}MnO_3$ compound is cooled in a static magnetic field through the Curie temperature $T_C \sim 70$ K, the magnetization hysteresis loops exhibit negative horizontal shifts (Figs. 1 – 3), which manifest the exchange bias phenomenon. The EB effect found is evident of an unidirectional exchange anisotropy interaction in the compound. The effect is associated with the boundaries of AFM and FM phases, which coexist at low temperatures in the spontaneously phase separated $Nd_{2/3}Ca_{1/3}MnO_3$ compound. In the magnetic field-cooled (FC) compound the interface exchange interaction drives the ferromagnetic clusters back to the original orientation when the magnetic field is removed. It leads to the exchange bias phenomena. For zero-field-cooling (ZFC) $H_{cool} = 0$ the exchange bias effect is absent (Figs. 1 – 3): the shift of the hysteresis loops is not seen when cooling in zero magnetic field at all the temperatures studied.

The magnetic field induced shift of the hysteresis loop was determined as $H_{EB} = (H_1 + H_2)/2$, the coercive field was defined as $H_C = (H_2 - H_1)/2$, where $H_1$ and $H_2$ are magnetic fields at which the magnetization is zero (Fig. 1, the insert). When the sample was cooled down to 10 K in magnetic field $H_{cool} = 1$ T, as the magnetic field was reduced from its maximum value, the magnetic moment becomes negative at $H_1 \sim -0.09$ T (decreasing branch of the loop), while on the increasing branch of the loop $M(H)$ changes sign at $H_2 \sim 0.04$ T. These two fields define the coercive field $H_c \sim 0.06$ T and the exchange bias field $H_{EB} \sim -0.03$ T.

We have found that the exchange bias effect and the coercive field in $Nd_{2/3}Ca_{1/3}MnO_3$ are strongly dependent on the temperature and cooling magnetic field. The examples of hysteresis loops at 10 K after zero-field-cooling and field-cooling in $H_{cool} = 5$ T, and at 35 K after zero-field-cooling and field cooling in $H_{cool} = 1$ T, are shown in Fig. 2 and Fig. 3, respectively. The EB effect was absent at 75 K for the all cooling fields studied (Fig. 1).

Temperature dependences of the exchange bias $H_{EB}$ and the coercive field $H_c$ are shown in Fig. 4. When temperature decreases the both $H_{EB}$ and $H_c$ absolute values are greatly growth, reaching their maximum value at low temperature. It indicates that below $T_C \sim 70$ K a FM contribution is present, which increases with decreasing temperature. The FM contribution can be associated with the FM clusters. The growth of the absolute value of $H_{EB}(T)$ applies that surface area between the AFM – FM coexisting phases increases due to the increase in the number of FM clusters. It is in a good agreement with our earlier neutron diffraction data: the FM component gradually growths with decreasing temperature below 70 K [22]. The EB effect, as well as the coercivity $H_c$, finally vanishes above $T_C \sim 70$ K, where there are no FM clusters, and AFM charge-ordered matrix with paramagnetic charge-disordered clusters coexist.

The competing magnetic interactions is known to lead to an exponential temperature dependent decay of $H_{EB}$ and $H_c$ in exchange biased systems [24]. Indeed, the temperature dependences of $H_{EB}$ and $H_c$ can be fitted by the phenomenological formula

$$H_{EB}(T) = H_{EB}^0 \cdot exp\left(-\frac{T}{T_1}\right)$$
$$H_C(T) = H_C^0 \cdot exp\left(-\frac{T}{T_2}\right) \quad , \tag{1}$$

where $H_{EB}^0$ and $H_C^0$ are the extrapolations of $H_{EB}$ and $H_c$ to the zero temperature; $T_1$ and $T_2$ are constants. We have got the following parameters $H_{EB}^0 = -44$ mT, $H_C^0 = 131$ mT, $T_1 = 20$ K, $T_2 = 15$ K. the fitting results (Fig. 4) give further support to the scenario that the EB in $Nd_{2/3}Ca_{1/3}MnO_3$ can be attributed to the competition between the FM and AFM interactions.

Cooling magnetic field dependence of the exchange bias $H_{EB}(H_{cool})$ and the coercive field $H_c(H_{cool})$ are shown in Fig. 5. The both dependences seems to be nonmonotonic. Absolute value of exchange bias effect decrease with cooling magnetic field growth, while the coercive field increase. It may be attributed to a nonmonotonic enlargement of ferromagnetic clusters in size with the growth of magnetic field. Presumably, low magnetic fields affect FM clusters and glasslike surface spins, orienting them along the magnetic field applied, while the applied magnetic fields as high as ~ 3 T lead to the enlargement of FM clusters in size at the expense of AFM background, which is in a good agreement with the threshold field of a field-induced AFM – FM transition $H_f$ ~ 3 T found in $Nd_{2/3}Ca_{1/3}MnO_3$ at low temperature [25].

It should be noted that the magnetic moment does not saturate in all the magnetic fields studied (up to 5 T). In magnetic fields 1 T > $H$ > 0.5 T, the dependence $M(H)$ is linear, and the increase in the magnetic moment may be attributed to the AFM phase surrounding FM clusters (Fig. 1). While the magnetic fields $H$ > 3 T lead to the growth of the FM phase fraction at the expense of AFM background. So, to found a response of the ferromagnetic phase in low magnetic fields, the linear AFM contribution ($T$=10 K, $H_{cool}$ = 1 T) was subtracted. The resulting magnetization $M_{FM}(H)$ associated with a contribution of FM clusters immersed in the charge-ordered AFM matrix, is shown in Fig. 6. The coercive field was defined as $H_c$ ~ 0.08 T, $H_{EB}$ ~ - 0.03 T. The magnetization $M_{FM}$ saturates in relatively low magnetic field about 0.4 T, which coincides with the magnetic field that effectively suppresses the glassy magnetic state of the compound [19]. The obtained saturation value $M_{FM}$ (1 T) ~ 0.45 $\mu_B$ (a contribution of $Nd^{+3}$ ions is included) well agrees with our previous neutron diffraction data [22].

In addition, it is need to note that a training effect which is a typical feature of EB systems, has been observed in our study as well. It is evident of a metastable equilibrium in the system, the data will be published elsewhere.

The results obtained show that exchange bias phenomena found in $Nd_{2/3}Ca_{1/3}MnO_3$ is tightly connected with its self-organized inhomogeneous magnetic state at low temperatures, representing coexisting AFM and FM domains. As the fraction and size of the FM clusters are sensitive to the temperature and cooling magnetic field, the EB effect may be tuned both by the temperature and cooling magnetic field.

**4. Conclusions**

The exchange bias behavior found in $Nd_{2/3}Ca_{1/3}MnO_3$ is induced by the interface exchange coupling between the FM clusters and AFM charge-ordered background; it is evident of the low temperature phase segregated state of the compound. The EB effect found is strongly temperature and cooling magnetic field dependent. It steady decreases with increasing temperature and vanishes above the Curie temperature. The EB effect nonmonotonously decreases with cooling magnetic field increase within $H = 0.2 - 5$ T range. The evolution of EB effect is associated with the evolution of the phase segregated state of the compound; it depends on the number and size of FM clusters embedded into AFM matrix at the temperature and magnetic field applied. The ferromagnetic moment $M_{FM}$ originating from the FM clusters has been found to saturate in a relatively low magnetic field about $H \sim 0.4$ T. The obtained saturation value $M_{FM}$ (1 T) $\sim 0.45$ $\mu_B$ is in a good agreement with our previous neutron diffraction data.

Authors are thankful to Dr. D. Khalyavin (ISIS, United Kingdom) for fruitful collaboration. We are thankful to Prof. M. F. Kharchenko and Prof. V. V. Eremenko for fruitful discussions. This work was supported by the Slovak Research and Development Agency under the contract No. APVV-0132-11 and by the State Fund of NAS of Ukraine under the contact No.4/11-H. One of the authors, E. F., is grateful to the SAIA (Slovak Academic Information Agency), for the award of a fellowship.

**Figure captions**

Fig. 1. (Color online) Hysteresis loops of $Nd_{2/3}Ca_{1/3}MnO_3$ at 10 K and at 75 K measured after zero-field-cooling and field-cooling ($H_{cool}$ = 1 T). Inset: enlarged view of the central region of the loop taken at 10 K.

Fig. 2. (Color online) Hysteresis loops of $Nd_{2/3}Ca_{1/3}MnO_3$ at 10 K measured after zero-field-cooling and field-cooling ($H_{cool}$ = 5 T). Inset: enlarged view of the central region of the loop.

Fig. 3. (Color online) Hysteresis loops of $Nd_{2/3}Ca_{1/3}MnO_3$ at 35 K measured after zero-field-cooling and field-cooling ($H_{cool}$ = 1 T). Inset: enlarged view of the central region of the loop.

Fig. 4. (Color online) Temperature dependence of the exchange bias field $H_{EB}$ and the coercive field $H_c$ for $Nd_{2/3}Ca_{1/3}MnO_3$ after field cooling ($H_{cool}$ = 1 T) and zero-field-cooling. Lines are the approximations $H_{EB}(T) = -44 exp\left(-\frac{T}{20}\right)$ and $H_C(T) = 131 exp\left(-\frac{T}{15}\right)$, correspondingly.

Fig. 5. Cooling field dependence of the exchange bias field $H_{EB}$ and the coercive field $H_c$ at 10 K for $Nd_{2/3}Ca_{1/3}MnO_3$ after field-cooling and zero-field-cooling. Lines are guides for eyes only.

Fig. 6. (Color online) Resulting magnetization $M_{FM}$ associated with magnetic contribution of FM phase, after subtracting the contribution of the AFM matrix ($T$ = 10 K, $H_{cool}$ = 1 T).

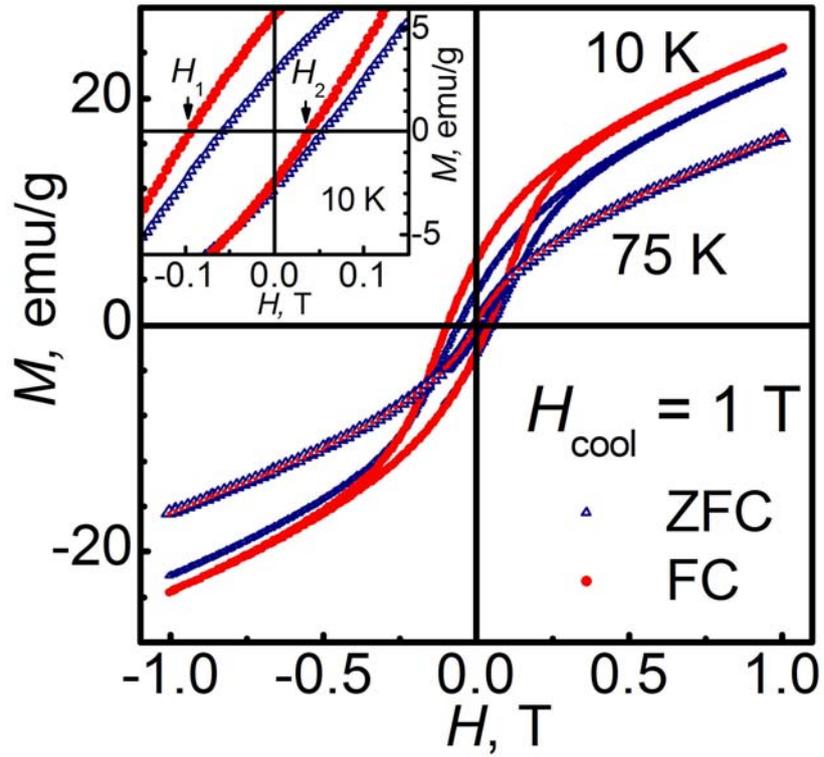

Fig. 1. (Color online) Hysteresis loops of $Nd_{2/3}Ca_{1/3}MnO_3$ at 10 K and at 75 K measured after zero-field-cooling and field-cooling ($H_{cool}$ = 1 T). Inset: enlarged view of the central region of the loop taken at 10 K.

E. Fertman, S. Dolya, V. Desnenko, M. Kajnakova, and A. Feher, Exchange bias associated with phase separation in $Nd_{2/3}Ca_{1/3}MnO_3$ manganite

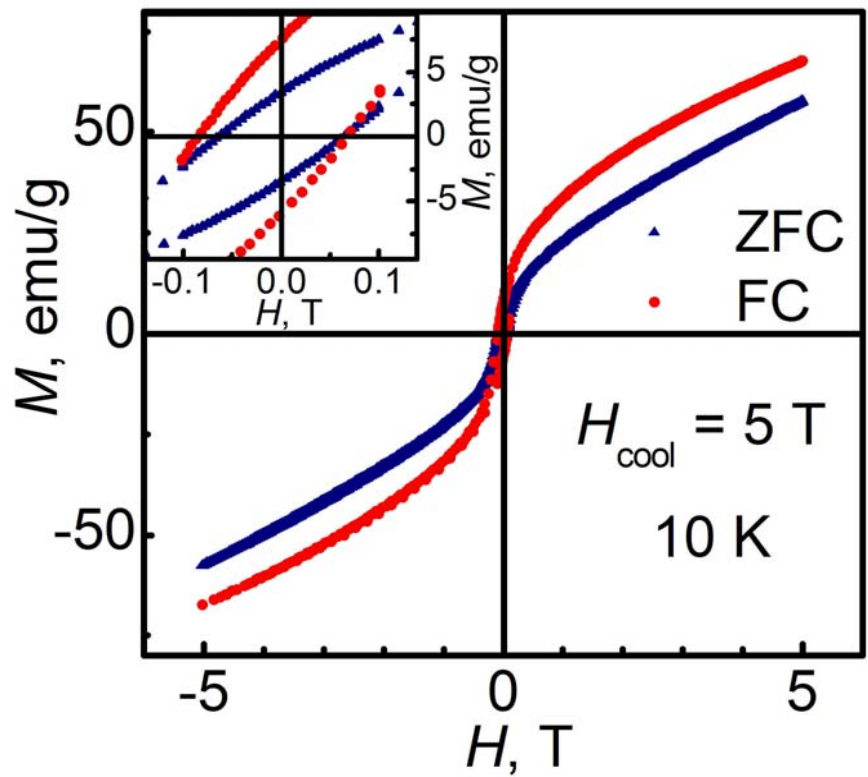

Fig. 2. (Color online) Hysteresis loops of Nd$_{2/3}$Ca$_{1/3}$MnO$_3$ at 10 K measured after zero-field-cooling and field-cooling ($H_{cool}$ = 5 T). Inset: enlarged view of the central region of the loop.

E. Fertman, S. Dolya, V. Desnenko, M. Kajnakova, and A. Feher, Exchange bias associated with phase separation in Nd$_{2/3}$Ca$_{1/3}$MnO$_3$ manganite

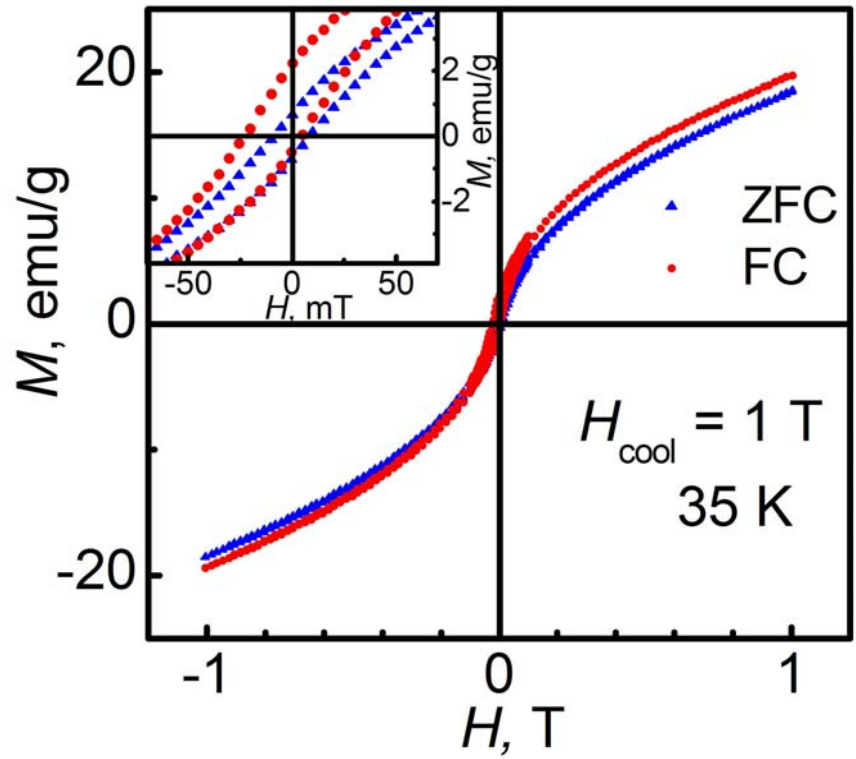

Fig. 3. (Color online) Hysteresis loops of $Nd_{2/3}Ca_{1/3}MnO_3$ at 35 K measured after zero-field-cooling and field-cooling ($H_{cool}$ = 1 T). Inset: enlarged view of the central region of the loop.

E. Fertman, S. Dolya, V. Desnenko, M. Kajnakova, and A. Feher, Exchange bias associated with phase separation in $Nd_{2/3}Ca_{1/3}MnO_3$ manganite

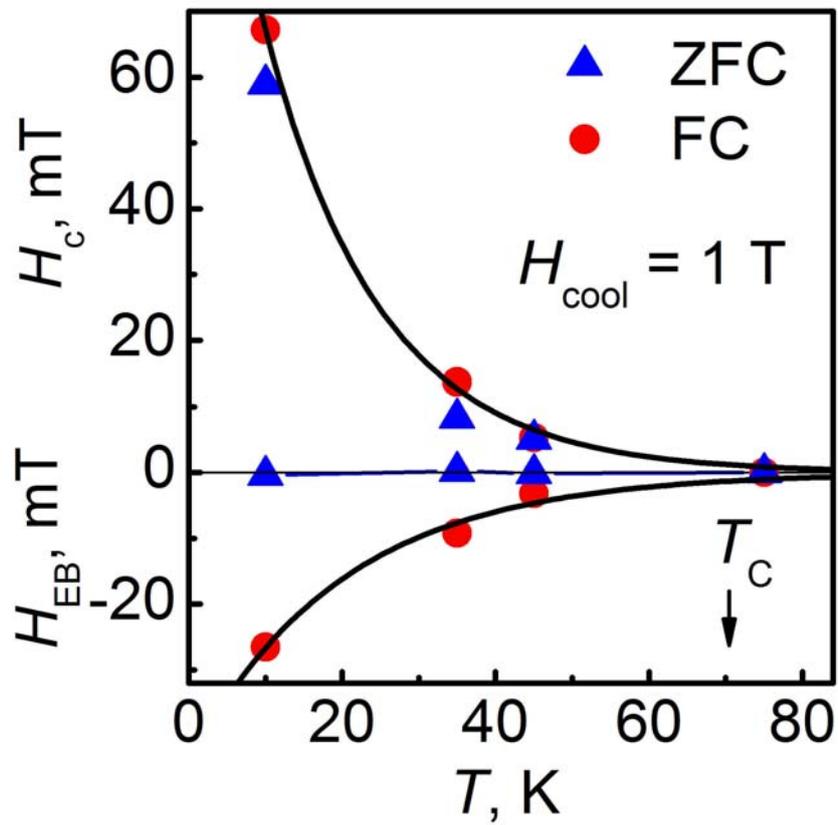

Fig. 4. (Color online) Temperature dependence of the exchange bias field $H_{EB}$ and the coercive field $H_c$ for $Nd_{2/3}Ca_{1/3}MnO_3$ after field cooling ($H_{cool}$ = 1 T) and zero-field-cooling. Lines are the approximations $H_{EB}(T) = -44 exp\left(-\frac{T}{20}\right)$ and $H_C(T) = 131 exp\left(-\frac{T}{15}\right)$, correspondingly.

E. Fertman, S. Dolya, V. Desnenko, M. Kajnakova, and A. Feher, Exchange bias associated with phase separation in $Nd_{2/3}Ca_{1/3}MnO_3$ manganite

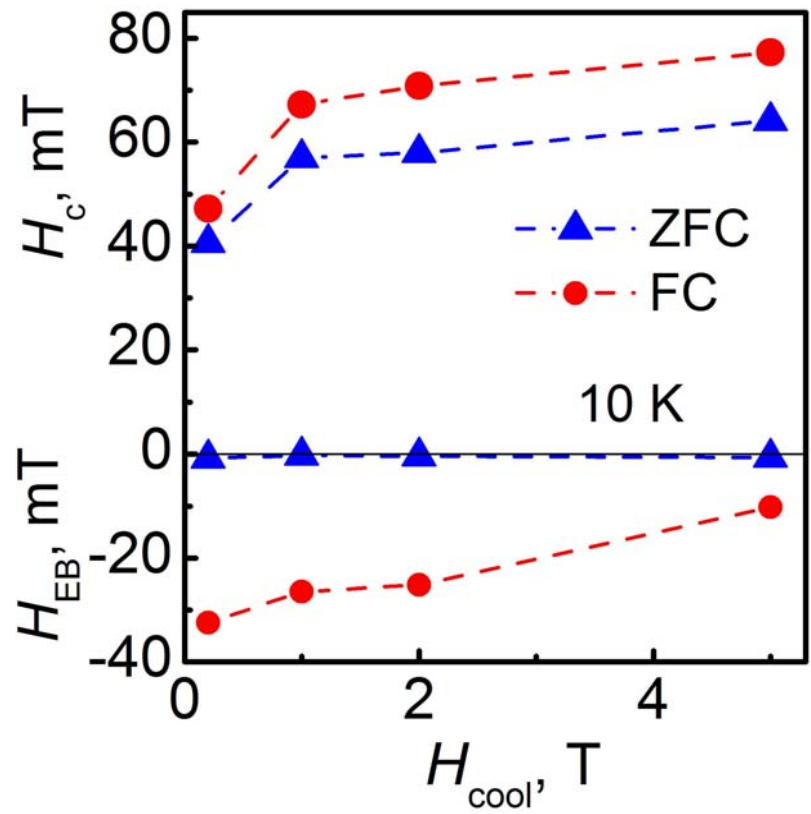

Fig. 5. Cooling field dependence of the exchange bias field $H_{EB}$ and the coercive field $H_c$ at 10 K for $Nd_{2/3}Ca_{1/3}MnO_3$ after field cooling and zero-field cooling. Lines are guides for eyes only.

E. Fertman, S. Dolya, V. Desnenko, M. Kajnakova, and A. Feher, Exchange bias associated with phase separation in $Nd_{2/3}Ca_{1/3}MnO_3$ manganite

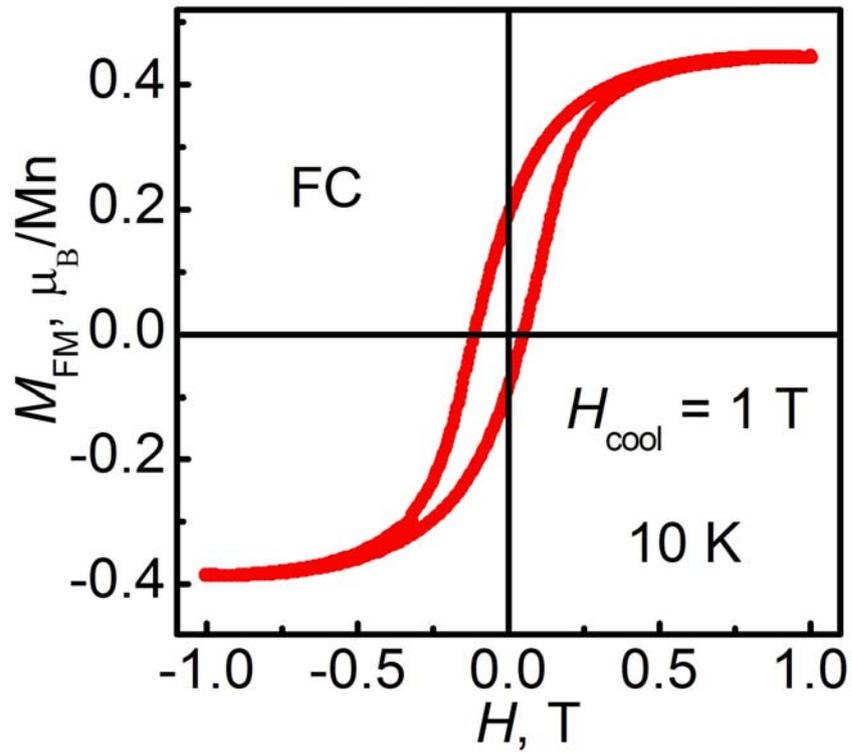

Fig. 6. (Color online) Resulting magnetization $M_{FM}$ associated with magnetic contribution of FM phase, after subtracting the contribution of the AFM matrix ($T$ = 10 K, $H_{cool}$ = 1 T).

E. Fertman, S. Dolya, V. Desnenko, M. Kajnakova, and A. Feher, Exchange bias associated with phase separation in $Nd_{2/3}Ca_{1/3}MnO_3$ manganite